# Nanoscale observation and control of quasiparticle induced magnetic noise in a superconducting resonator

Senlei Li[1], Shane P. Kelly[2], Jingcheng Zhou[1], Hanyi Lu[3], Yaroslav Tserkovnyak[2], Hailong Wang[1,*], and Chunhui Rita Du[1,*]

[1]School of Physics, Georgia Institute of Technology, Atlanta, Georgia 30332, USA
[2]Department of Physics and Astronomy, University of California, Los Angeles, California 90095, USA
[3]Department of Physics, University of California, San Diego, La Jolla, California 92093, USA

[*]Corresponding authors: hwang3021@gatech.edu; cdu71@gatech.edu

**Abstract**: Superconducting circuits are arguably taking a leading role in driving the ongoing quantum technological revolution. A detailed knowledge of the microscopic fluctuating electromagnetic properties plays an important role in advancing the circuitry design, testing, and material integration of cutting-edge superconducting quantum electronics. Here we report scanning nitrogen-vacancy (NV) quantum sensing of local magnetic noise environment of an on-chip superconducting resonator. We find that quasiparticle-induced fluctuating magnetic fields can drive NV spin relaxation, which shows a peak value around the superconducting transition point of niobium at the thermal equilibrium state. External microwave driving at the resonator mode frequency significantly increases the quasiparticle density, leading to enhancement of magnetic noise. We further perform optically detected magnetic resonance measurements to demonstrate quasiparticle magnetic noise mediated off-resonant dipole coupling between the NV center and niobium resonator. Our work reports experimental observation of the Hebel-Slichter peak signature by an external sensor outside of a superconductor. The presented study also highlights the advantages of quantum sensors in investigating miniaturized superconducting devices, providing insights into their future performance improvements.

The past decade has witnessed waves of quantum breakthroughs to transform cutting-edge technological development and scientific research [1–5]. A grand strategy of this new era is to harness quantum mechanics principles such as coherence, superposition, and strong coupling to develop next-generation sensing, computing, simulation, and communication technologies endowed with previously inaccessible merits and functionalities [4,6,7]. Superconducting circuits naturally stand out as a promising building block of the ongoing quantum revolution [1,8]. In comparison with other potential quantum operational platforms, the superconducting ones show clear advantages in solid-state scalability, electromagnetic tunability, and strong coupling with external fields, and are arguably taking a leading role in advancing the ongoing quantum innovation [1,8,9]. To date, a "prototype" of superconducting quantum computers has been experimentally demonstrated [1,8], and state-of-the-art quantum simulators [10], quantum registers [11], and quantum networks [12,13] built on superconducting devices are underway.

Despite the significant progress and enormous promise, technical challenges remain to be resolved in order to push the current superconducting quantum initiative to the next level [8,14]. At the present technological level, design, testing, and evaluation of the performance and functionalities of superconducting devices and integrated circuits mainly rely on theoretical simulations, global microwave and electrical transport characterizations. A comprehensive knowledge of local properties, especially the extrinsic and intrinsic electromagnetic fluctuations, of superconducting circuits remains elusive, which hinders further improvement of their scalability, quality-factor, and quantum coherence merit [14,15]. In fact, many of the physical "defects" and electromagnetic "impurities" resulting in quantum decoherence, energy dissipation, or unwanted malfunctions occur over the micrometer or even nanometer length scale, which are difficult to access by conventional sensing approaches [14].

Here, we introduce scanning nitrogen-vacancy (NV) microscopy [16–23] to timely address this challenge. We report multimodal quantum sensing of local fluctuating electromagnetic field environment of an on-chip niobium (Nb) superconducting resonator [24–28]. Taking advantage of NV spin relaxometry, we find that quasiparticle-induced magnetic noise emanating from the Nb resonator shows a similar Hebel-Slichter peak [29] feature around the superconducting transition point at the thermal equilibrium state. External microwave driving at the resonator mode frequency significantly increases the quasiparticle density, leading to enhancement of the magnetic noise emanating from Nb. We further perform optically detected magnetic resonance (ODMR) measurements, demonstrating quasiparticle magnetic noise mediated off-resonant NV detection of the superconducting resonator mode. Our work highlights the advantages of quantum sensors on investigating nanoscale electromagnetic properties of miniaturized superconducting electronics, bringing valuable insights into future circuitry design and improvements of superconducting quantum devices [8]. The observed off-resonant NV detection of superconducting resonator mode also opens new pathways for developing integrated hybrid systems to advance the burgeoning quantum innovation [30–34].

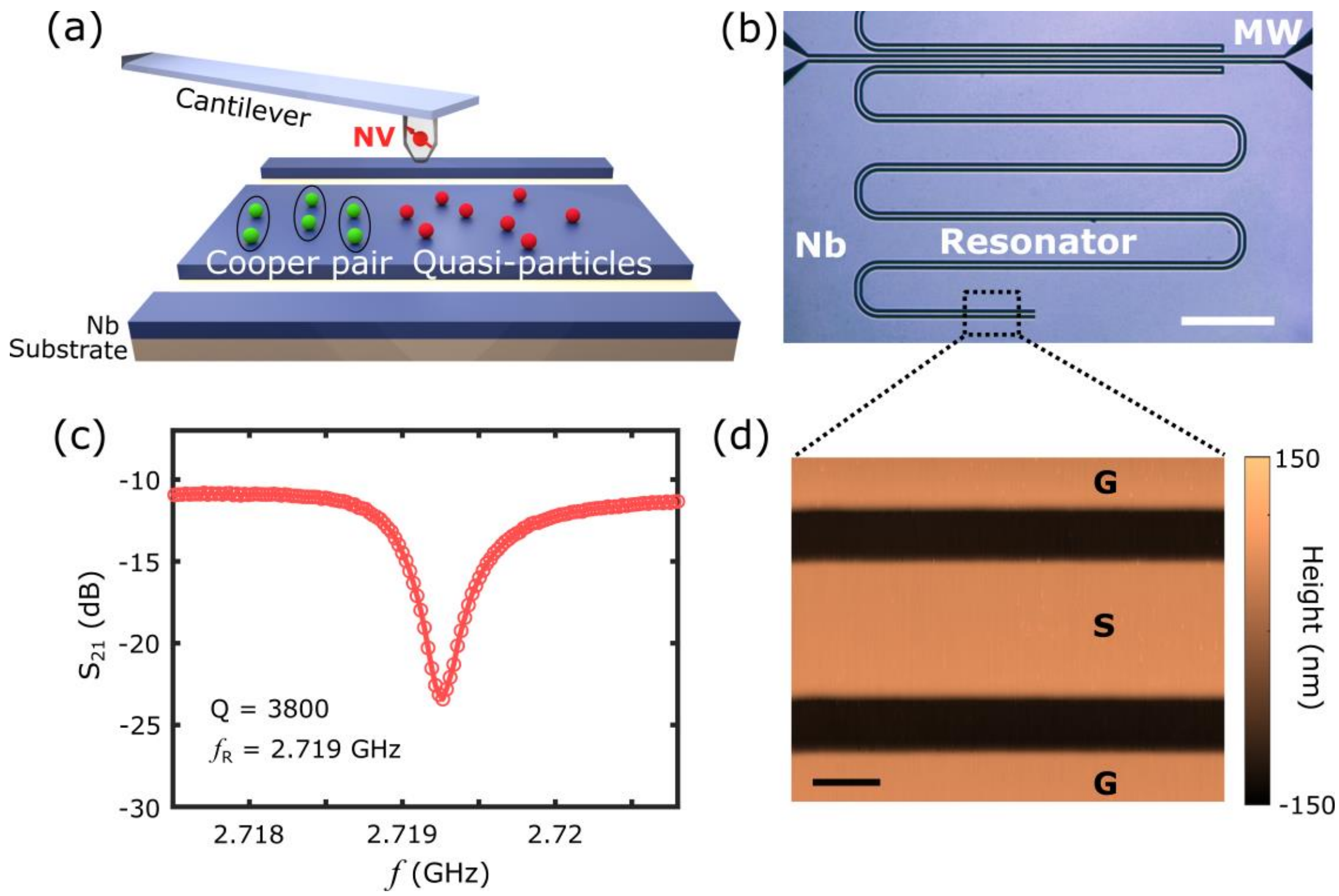

**FIG. 1.** Scanning quantum sensing of an on-chip superconducting resonator. (a) Schematic illustration of scanning NV sensing of quasiparticle magnetic noise from the coplanar signal line of an on-chip Nb resonator. (b) Optical microscopy image of a patterned Nb resonator device. Scale bar is 500 μm. (c) Microwave transmission of the superconducting resonator mode measured at 2 K without external magnetic field. (d) Atomic force microscopy imaging of topography of the Nb coplanar waveguide. Scale bar is 10 μm.

Before presenting experimental results, we first discuss the detailed measurement platform and device structure. We utilize an NV center implanted at the end of a nanopillar possessed by a diamond cantilever [35] to perform scanning-probe-based single-spin quantum sensing of an on-chip resonator as illustrated in Fig. 1(a). NV sensing takes advantage of the quantum mechanical nature of an isolated $S$ = 1 electron spin to achieve ultrasensitive detection of weak static and dynamic magnetic fields [17,36]. Figure 1(b) shows an optical microscopy image of a lithographically defined Nb superconducting circuit, which consists of a coplanar microwave feedline capacitively coupled with a resonator patterned on a sapphire substrate (see Supplemental Material [37] section 1 for details). The length of the signal line is 12 mm with one-end open-circuited, resulting in a characteristic resonator frequency of ~2.7 GHz. We characterize the microwave transmission performance of the Nb resonator through the feedline using a vector network analyzer. Figure 1(c) presents a resonator mode spectrum measured at 2 K with zero external magnetic field. It shows a microwave absorption dip at 2.719 GHz with a full width half maximum linewidth of ~0.7 MHz, corresponding to a quality factor of ~3,800. In the current study, our quantum sensing measurements mainly focus on the device area close to the shorted end of the signal line [Fig. 1(d)], where the amplitude of the microwave magnetic field generated by the Nb resonator is expected to reach the maximum for the convenience of NV spin control.

We now discuss NV quantum sensing of local fluctuating magnetic fields emanating from the Nb resonator. For a clean superconducting system such as Nb, quasiparticles featuring elementary electron and hole excitations from broken Cooper pairs constitute one of the major sources of electromagnetic noise as illustrated in Fig. 2(a) [29,38]. Figure 2(b) shows a sketch of 2D quasiparticle dispersion and occupation with a minimum energy of the *s*-wave superconducting energy gap. According to the Bardeen–Cooper–Schrieffer (BCS) theory, spin excitations in a superconductor with a singlet-spin pairing involve energy transfer between a pair of quasiparticles

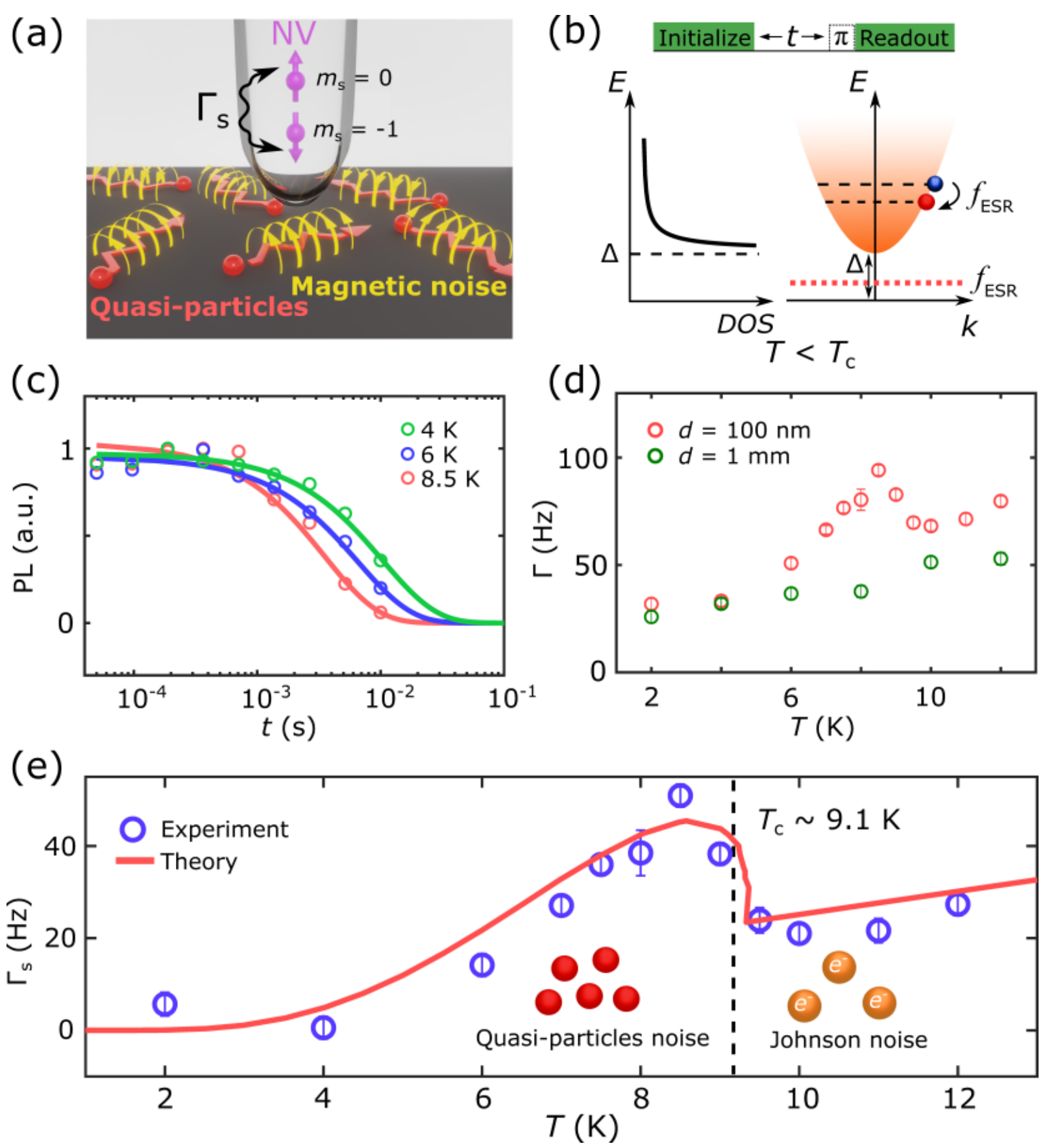

**FIG. 2.** Quantum sensing of quasiparticle-generated magnetic noise. (a) Thermally driven random motions of quasiparticles produce fluctuating magnetic fields that can be detected by an NV center in a diamond cantilever. (b) Sketch of 2D dispersion and density of states (DOS) of quasiparticles in a *s*-wave superconductor. The occupation function follows the Fermi-Dirac distribution as indicated by the fading colors. "Two-quasiparticle" scattering induced magnetic noise at ESR frequencies drives NV spin relaxation. Top right: pulsed optical and microwave sequences for NV spin relaxometry measurements. The $\pi$ pulse at the corresponding NV ESR frequency is used to invert the spin population prior to NV readout. The pulse duration is ~100 ns, orders of magnitude smaller than the NV $T_1$ time. Thus, the $\pi$ pulse does not drive the superconductor out of equilibrium. (c) A series of NV spin relaxometry spectra recorded at 4 K, 6 K, and 8.5 K. Variations of NV photoluminescence (PL) measured as a function of the delay time *t*. (d) Temperature dependence of NV spin relaxation rate Γ measured when the NV center is positioned 100 nm and 1 mm above the Nb sample. (e) Superconductivity-induced NV spin relaxation rate $\Gamma_s$ as a function of temperature between 2 K and 12 K. $\Gamma_s$ is driven by fluctuating magnetic fields generated by quasiparticles or Johnson–Nyquist noise from electrons below and above the $T_c$ of Nb, respectively.

[38]. A quasiparticle with a frequency $f + \Delta f$ can be scattered to another state with a frequency $f$, leading to emission of magnetic noise at a frequency $\Delta f$. When $\Delta f$ matches the NV electron spin resonance (ESR) frequency $f_{\mathrm{ESR}}$, quasiparticle induced magnetic noise will drive NV spin relaxation from $m_s = 0$ to $m_s = \pm 1$ states, resulting in enhanced NV relaxation rate that is proportional to the magnitude of local fluctuating magnetic fields transverse to the NV spin axis [38].

Here, we utilize NV spin relaxometry (see Supplemental Material section 2 for details) to experimentally evaluate the quasiparticle-induced electromagnetic noise emanating from the on-chip superconducting resonator [21,35,39,40]. Our measurements start from the NV-to-sample distance of ~100 nm to ensure sufficiently high field sensitivity. Figure 2(c) presents a series of NV spin relaxation spectra recorded at different temperatures. By measuring the spin-dependent NV photoluminescence (PL) as a function of the delay time and fitting it to a theoretical model, the occupation probabilities of NV spin states can be quantitatively obtained, allowing for extraction of NV spin relaxation rate $\Gamma$ as summarized in Fig. 2(d) (see Supplemental Material section 2 for details). When retracting the diamond cantilever ~1 mm away from the surface of the Nb sample, we further perform control measurement to characterize the intrinsic NV spin relaxation rate. By subtracting the linearly interpolated intrinsic contribution, superconductivity-induced NV relaxation rate $\Gamma_s$ can be deduced as presented in Fig. 2(e). One can see that $\Gamma_s$ shows a peak value at a temperature just below the superconducting transition point $T_c$ of Nb, which resembles the Hebel-Slichter peak observed in nuclear spin relaxation measurements of $s$-wave superconductors [29]. The major difference here is that the NV spin probes magnetic noise outside a superconductor, while the nuclear spin in nuclear magnetic resonance (NMR) measurement detects local hyperfine noise inside a sample [29,38].

We have developed a generic theory based on the BCS model to rationalize well the observed temperature dependent variations of quasiparticle-induced magnetic noise in Nb in the thermal equilibrium state (see Supplemental Material section 3 for details) [38]. For a two-dimensional $s$-wave superconductor, the NV spin relaxation rate $\Gamma_s$ related to current fluctuations associated with quasiparticles can be expressed as follows [38]:

$$\Gamma_{\mathrm{s}} = \left(\frac{\gamma}{2}\right)^2 \int d^2\boldsymbol{k}\, \frac{e^{-k2d}}{c^2} C_{\mathrm{JJ}}(\boldsymbol{k}, f_{\mathrm{ESR}}) \tag{1}$$

where $C_{\mathrm{JJ}}$ is the transverse-current spectral density in Nb, $\gamma$ is the gyromagnetic ratio, $\boldsymbol{k}$ represents the wave vector of quasiparticles, $c$ is the speed of light, and $d$ characterizes the NV-to-sample distance. The red curve in Fig. 2(e) plots the theoretically predicted temperature dependence of $\Gamma_s$ over the superconducting phase transition point ($T_c$) of Nb. Note that our model assumes electronic noise in Nb increases linearly with the measurement temperature when above the $T_c$ and the presented calculations are normalized by the experimental data at 9.5 K (see Supplemental Material section 3 for details). Invoking an intuitive physical picture, the magnitude of fluctuating magnetic noise from superconducting Nb depends on the density of quasiparticle states which has a square root singularity $DOS(E) \propto 1/\sqrt{E^2 - \Delta(T)^2}$, where the superconducting energy gap $\Delta$ decreases with increasing temperature and vanishes at $T_c$ [29,38,41]. This singularity behavior results in the low-frequency magnetic noise probed by the NV center ( $f_{\mathrm{ESR}} < \Delta$, $h$ is the Planck constant) reaching a peak value around the $T_c$ [38]. When the measurement temperature is well below $T_c$, thermally induced electromagnetic fluctuations decrease exponentially due to the reduced quasiparticle density in Nb [38]. Electromagnetic fluctuations remain active in Nb above $T_c$ owing to free electron motion-driven Johnson-Nyquist noise in the normal metal phase [39].

The satisfactory agreement between our theory and experimental results confirms the measured NV spin relaxation is indeed driven by quasiparticle magnetic noise in superconducting Nb. Other potential mechanisms of magnetic noise such as critical fluctuations are expected to be only active in a very narrow temperature window near the $T_c$ due to strong spatial overlap of Cooper pairs in Nb, which is experimentally challenging to access [42]. This is different from high temperature cuprate superconductors, whose critical fluctuations are stronger and more accessible to be detected [42–44].

In addition to temperature, microwave magnetic fields generated by the on-chip resonator provide an alternative tuning knob to control quasiparticle excitation in superconducting Nb. When the Nb resonator is driven to the resonant state, quasiparticles will absorb microwave photons and are excited to higher energy levels as illustrated in Fig. 3(a) [45,46]. This process will unlock additional Cooper pairs breaking in superconducting Nb in comparison with the thermal equilibrium case, leading to an increased quasiparticle density and larger emanating magnetic noise at NV ESR frequencies [Fig. 3(b)] [46]. To investigate the microwave control of the quasiparticle magnetic noise, we measure the NV spin relaxation rate with external microwave driving using the pulse sequence shown in the top panel of Fig. 3(b). Here, microwave currents with a variable frequency $f$ are applied through the coplanar microwave feedline to control microwave photon generation in the superconducting Nb resonator. Figure 3(c) presents the NV relaxation rate $\Gamma_s$ measured as a function of $f$. One can see that $\Gamma_s$ reaches a maximum when $f$ matches the resonator frequency $f_R$, which is expected due to the most efficient microwave photon generation under the resonant condition. When $f = f_{\mathrm{R}}$, $\Gamma_{\mathrm{s}}$ exhibits a characteristic polynomial increasing behavior as a function of the input microwave power owing to the nonlinear photon-quasiparticle interaction as shown in Fig. 3(d) [45]. We add that the presented NV spin relaxometry measurements are peformed at $f_{\mathrm{ESR}}$ = 2.865 GHz, well above $f_{\mathrm{R}}$ (~2.7 GHz). Thus, the enhanced NV spin relaxation rate observed is dominated by resonance induced increase of quasiparticle density in superconducting Nb rather than the direct driving/decoherence of the NV center by microwave fields.

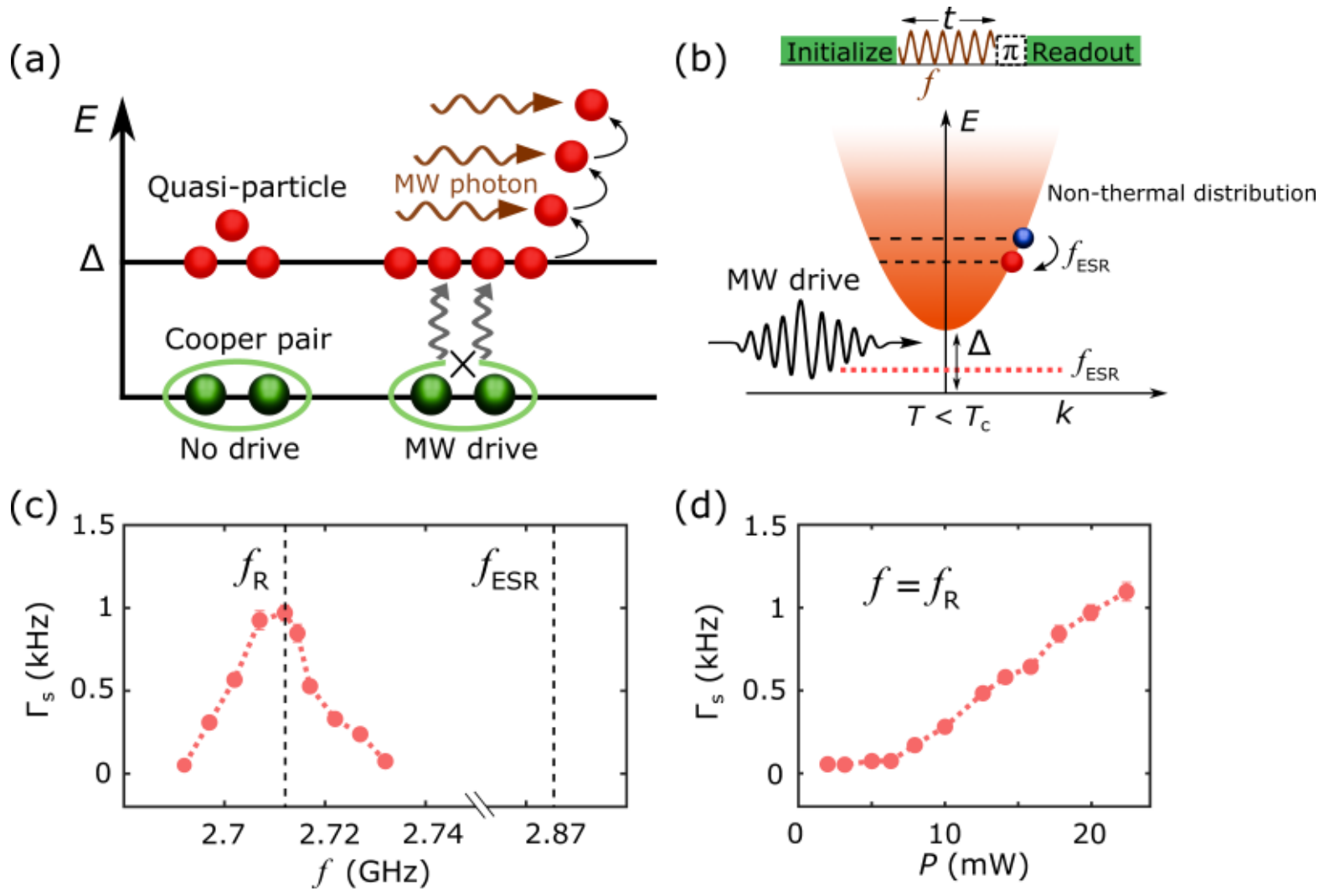

**FIG. 3.** Microwave control of quasiparticle-generated magnetic noise. (a) Schematic of microwave photon assisted quasiparticle generation in superconducting Nb. (b) Microwave excitation at the resonator frequency increases the number of quasiparticles and emanating magnetic noise at the NV ESR frequency. Top: pulsed optical and microwave sequences for NV relaxometry measurements under external microwave driving of the Nb resonator. (c) NV spin relaxation rate $\Gamma_s$ shows a peak value when the frequency $f$ of external driving microwave field matches the frequency $f_R$ of the Nb resonator mode. The input external microwave power $P$ is 20 mW. (d) $\Gamma_s$ measured as a function of microwave power under the resonant condition ($f = f_R$). The measurement temperature is 6 K, and the NV-to-sample distance is ~100 nm for results presented in Figs. 3(c)-3(d).

The microwave tunable quasiparticle magnetic noise generation opens a new pathway to establish dynamic coupling between NV centers and superconducting resonators. Lastly, we present ODMR measurement results (see Supplemental Material section 4 for details) [36,44] to demonstrate quasiparticle scattering mediated off-resonant NV detection of the Nb resonator mode. An ODMR map presented in Fig. 4(a) plots the normalized NV PL intensity as a function of the applied microwave frequency $f$ and external magnetic field $B_{\mathrm{ext}}$ along the NV spin direction at 6 K. Interestingly, we observe two curved lines of $f_+$ and $f_-$ emerging at ~2.88 GHz, which is due to the decrease of NV fluorescence when the external microwave drive frequency $f$ matches the NV ESR conditions [17]. Here, $f_+$ and $f_-$ represent the ESR frequencies of NV spin transitions between the $m_s = 0$ and $m_s = \pm 1$ states. Deviation from the conventional straight ESR lines is attributed to the Meissner effect of superconducting Nb [47], which partially shields the effective magnetic field detected by the NV center. Notably, NV PL intensity also decreases when $f$ matches the resonator frequency $f_{\mathrm{R}}$. The off-resonant coupling effect becomes significantly pronounced when $f_{\mathrm{R}}$ approaches and eventually matches the NV ESR frequency. Figure 4(b) shows a linecut of the ODMR spectrum at $B_{\mathrm{ext}} = 45$ G, which shows clear signatures of NV ESR and the Nb resonator mode. As discussed above, the presented ODMR detection of Nb resonator mode is fundamentally driven by the microwave photon assisted quasiparticle generation in superconducting Nb [46]. At the resonant state, the increased quasiparticle number naturally enhances magnetic noise at frequencies $f_\pm$ that will drive NV spin relaxation and variations of PL intensity as shown in Figs. 4(a)-4(b).

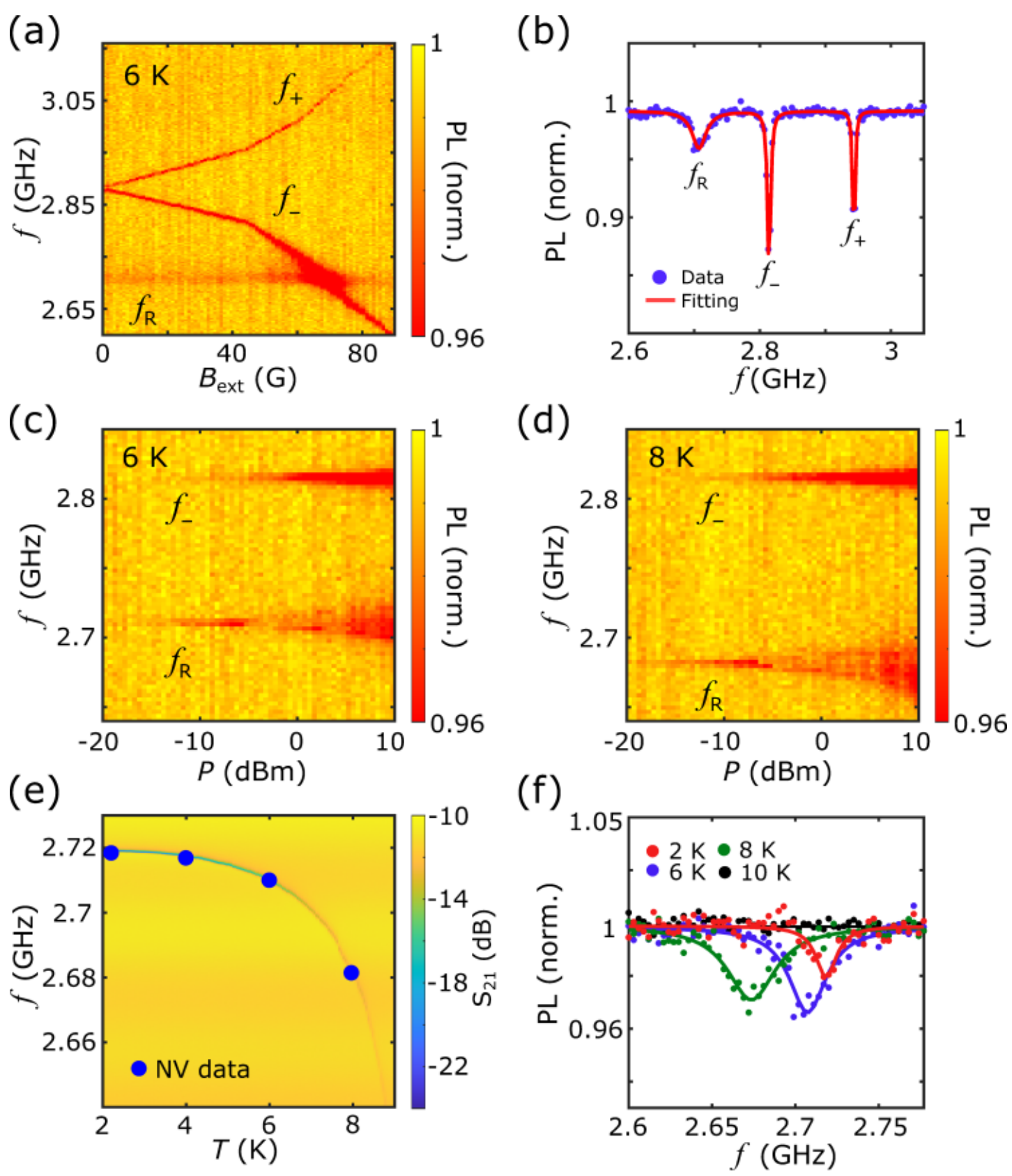

**FIG. 4.** Off-resonant NV ODMR detection of the superconducting resonator mode. (a) ODMR map shows quasiparticle noise mediated off-resonant detection of the Nb resonator mode at 6 K. (b) A linecut of the ODMR map at $B_{\mathrm{ext}}$ = 45 G highlights the NV ESR and Nb resonator mode. (c)-(d) Linecuts at $B_{\mathrm{ext}}$ = 45 G of ODMR maps measured as a function of input microwave power from −20 dBm to 10 dBm at 6 K and 8 K, respectively. (e) Microwave transmission coefficient $S_{21}$ of the on-chip Nb resonator measured as a function of frequency $f$ and temperature $T$. The blue points represent the resonant frequency obtained from NV ODMR measurements. (f) A zoomed-in view of NV ODMR detected Nb resonator mode at 2 K, 6 K, 8 K, and 10 K. Points are the experimental raw data and solid lines represent fitting curves. The NV-to-sample distance is ~500 nm for presented NV ODMR measurement results.

The presented NV ODMR measurements provide an accessible way to optically characterize the frequency, amplitude, and quality factor of the superconducting Nb resonator mode. Figures 4(c)-4(d) present linecuts at $B_{\mathrm{ext}} = 45$ G of measured NV ODMR maps as a function of input microwave power at 6 K and 8 K, respectively. The Nb resonator mode is clearly observed and the measured $f_{\mathrm{R}}$ shows a slight decrease when temperature increases from 6 K to 8 K. The optically detected Nb resonant frequency agrees well with the conventional microwave measurement results as summarized in Fig. 4(e). We can see that $f_{\mathrm{R}}$ gradually shifts towards the low-frequency end when the temperature increases due to variations of kinetic inductance of the Nb resonator [48]. Figure 4(f) plots a zoomed-in view of ODMR detected Nb resonator mode at temperatures from 2 K to 10 K. It is instructive to note that the measured linewidth of the superconducting resonator mode increases with temperature, consistent with the concurrent decrease of the resonator quality factor [48]. One can also see that the amplitude of the resonator mode shows a non-monotonic variation with the temperature. The initial increase from 2 K to 6 K is attributed to the enhanced quasiparticle density in Nb when the system temperature approaches the superconducting phase transition point $T_{\mathrm{c}}$. The following reduction of the resonant mode strength in the temperature window from 6 K to 10 K is dominated by the decay of the microwave transmission performance of the Nb resonator [48]. The resonator mode eventually disappears, as expected, in the NV ODMR map when the measurement temperature is above $T_{\mathrm{c}}$ (see Supplemental Material section 4 for details).

In summary, we have demonstrated multimodal scanning-probe quantum sensing of local magnetic noise environment of an on-chip superconducting resonator. We observe quasiparticle-induced fluctuating magnetic fields via both NV spin relaxometry and ODMR measurements. The measured NV spin relaxation rate shows a peak value around the $T_{\mathrm{c}}$ of Nb at the thermal equilibrium state, revealing a similar Hebel-Slichter peak [29] feature in *s*-wave superconductors. We show that external microwave driving can effectively increase the quasiparticle density under the resonant condition, leading to enhanced magnetic noise emanating from the Nb resonator. Because superconducting resonators are utilized to readout and couple qubit states in superconducting quantum computers [49–51], we share the optimism that the presented scanning NV microscopy platform can be extended readily to resolve spatially dependent electromagnetic noise induced by microscopic vortices, defects and structures in superconducting resonators, providing valuable insights into future circuitry design to reduce unwarranted quantum decoherence in superconducting qubit operations [14,46,52]. The cryogenic scanning NV quantum metrology techniques presented in the current work will also bring up new opportunities for evaluating nanoscale electromagnetic behaviors and performance of solid-state superconducting devices (see Supplemental Material section 5 for details), advancing the next-generation quantum electronics innovation [43,53–55].

**Acknowledgements**. The authors are grateful to Zelong Xiong and Mengqi Huang for help on device preparation and characterizations. The quantum sensing measurements are supported by the Office of Naval Research (ONR) under grant No. N00014-23-1-2146. The hardware development of scanning NV microscopy is supported by the U.S. Department of Energy (DOE), Office of Science, Basic Energy Sciences (BES), under award No. DE-SC0024870. Device fabrication is supported by the Air Force Office of Scientific Research (AFOSR) under award No. FA9550-21-1-0125. The software development of scanning NV magnetometry is supported by the Alfred. P. Sloan Foundation (FG-2024-21387). The work at UCLA is supported by U.S. National Science Foundation (NSF) under grant No. DMR-2049979.